\documentclass[twocolumn,english,prl]{revtex4-1}
\usepackage[T1]{fontenc}
\usepackage[latin9]{inputenc}
\usepackage{color}
\usepackage{babel}
\usepackage{amsmath}
\usepackage{amssymb}
\usepackage{graphicx}
\usepackage{multirow}
\usepackage{float}
\usepackage[unicode=true,
 bookmarks=false,
 breaklinks=false,pdfborder={0 0 0},pdfborderstyle={},backref=false,colorlinks=true]
 {hyperref}
\hypersetup{
 colorlinks}

\makeatletter
\usepackage{babel}

\makeatother

\begin{document}
\title{Enhancement of Microwave to Optical Spin-Based Quantum Transduction
via a Magnon Mode}
\author{Tharnier O. Puel$^{1}$, Adam T. Turflinger$^{2}$, Sebastian P. Horvath$^{2}$,
Jeff D. Thompson$^{2}$, Michael E. Flatt\' e$^{1,3}$\\
\textit{\small{}$^{1}$Department of Physics and Astronomy, University
of Iowa, IA 52242, USA}\\
\textit{\small{}$^{2}$Department of Electrical and Computer Engineering, Princeton
University, NJ 08544, USA}\\
\textit{\small{}$^{3}$Department of Applied Physics, Eindhoven University
of Technology, Eindhoven, The Netherlands}}
\begin{abstract}
We propose a new method for converting single microwave photons to single optical
sideband photons based on spinful impurities
in magnetic materials. This hybrid system is advantageous over previous proposals because 
(i) the implementation allows much higher transduction rates ($10^3$ times faster at the same optical pump Rabi frequency) than state-of the art devices, 
(ii) high-efficiency transduction is found to happen
in a significantly larger space of device parameters (in particular, over $1 \text{ GHz}$ microwave detuning), 
and (iii) it does not require mode volume {\color{black}matching} between optical and microwave resonators. 
We identify the needed magnetic interactions as well as potential materials systems to enable this speed-up using erbium dopants for telecom compatibility. This is an important step towards realizing high-fidelity entangling operations between remote qubits {\color{black} and will provide additional control of the transduction through perturbation of the magnet}.
\end{abstract}
\maketitle


Coherent transduction of microwave photons to optical photons\cite{Xiang-2013} {\color{black}will enable} scaling of dilution-fridge ({\it e.g.} superconducting\cite{Arute:2019aa}) quantum processors via remote entangling operations over room-temperature optical fiber. 
Demonstrated high-efficiency quantum transduction platforms include optomechanical transducers, \citep{Bagci:2014aa,Andrews:2014aa,Bochmann:2013aa}, the electro-optic effect in
lithium niobate \citep{PhysRevA.81.063837,Rueda:16}, the magneto-optic effect in $\text{Y}_3\text{Fe}_5\text{O}_{12}$ (YIG)~\cite{ghirri2023ultra},  Rydberg atom clouds\citep{PhysRevA.85.020302,PhysRevA.89.010301,PhysRevLett.120.093201,Tu:2022aa,Kumar-2023}, and dilute ensembles of rare-earth ions in a crystal \citep{PhysRevLett.113.203601,PhysRevA.92.062313,PhysRevA.100.033807,Bartholomew:2020aa,Rochman:2023aa,2024yvo4transducer}.  Optomechanical transduction enabled the first optical
readout of Rabi oscillations in a superconducting qubit \citep{Mirhosseini:2020aa,Delaney:2022aa}, lithium niobate transducers have achieved moderate conversion efficiencies with the introduction of only a small amount of conversion noise and MHz bandwidth\citep{doi:10.1126/sciadv.aar4994,Sahu:2022aa}, Rydberg atom transducers have demonstrated  near-unit-efficiency frequency conversion between microwave and optical photons with high
bandwidth, and rare-earth ions (particularly using Er for telecom wavelength compatibility) have demonstrated coherent chip-integrated  wavelength transducers.
However, each of these approaches faces critical challenges \textcolor{black}{in achieving high efficiency,  bandwidth well-matched to the $10\text{-}100$ ns timescale for gate operations in superconducting qubits~\cite{Place-2021}, and integration with superconducting qubits in the same device. The optical pump necessary to bridge the frequency gap between microwave and optical photons specifically creates notable technical issues including heating \cite{Azuma:2015aa} and destruction of cooper pairs in superconducting circuits \cite{PhysRevApplied.21.014022} leading to noise and reduced efficiencies.
Rare-earth ion transducers have produced high efficiencies at MHz-scale bandwidths  \cite{2024yvo4transducer}, but scaling to repetition rates beyond the kHz-level with quantum-limited photon noise will require higher coupling between the ensemble and microwave photons as well as the mitigation of heating from the optical pump.}

Direct transduction using  magnets  has been proposed before on account of their strong coupling to microwave photons (e.g. in YIG\cite{ghirri2023ultra,PhysRevB.101.214414} and $\text{GdVO}_4$\citep{PhysRevB.101.214414}). \textcolor{black}{However, these crystals lack the sharp optical lines of rare-earth ions, so the transducers are limited by the weak Brillouin scattering interaction between optical photons and magnons in spite of the high microwave photon couplings} \citep{Liu-2016,PhysRevLett.121.199901,Zhu:20,PhysRevApplied.18.024046}.
Alternatively, fully-concentrated rare-earth crystals leverage the strong coupling of microwave photons to magnons
and the optical transitions of rare-earth ions\citep{PhysRevA.99.063830}, but demonstrating frequency conversion has been a challenge on account of the very high optical density of the fully concentrated rare-earth crystals\citep{berrington2022negative}.

\textcolor{black}{We propose to overcome the weak optical interaction of magnets using dilute doping of an optically active defect into a magnetic material, maintaining the strong microwave-magnon coupling while enhancing the optical coupling via the defects.} This proposal relies on attainable narrow microwave linewidths in magnets~\cite{Levy-1999,doi.org/10.1002/pssb.201900644,ASSIS2023170388,Zollitsch-2023} and optical linewidths in rare-earth ensembles~\cite{PhysRevB.73.075101,Zhong-2015,Kukharchyk_2018,doi:10.1021/acs.nanolett.9b03831,LAFITTEHOUSSAT2022100153,PhysRevB.105.224106}, 
and we effectively reduce the problem to creating an appropriate interaction between the magnet and dopants within several proposed materials. 
We demonstrate how the strengthened coupling, to both single microwave and optical photons, enables transducers with significantly increased effective transduction rates (as much as $10^3$ times faster at the same optical pump Rabi frequency) and bandwidths ($\sim 1 \text{ GHz}$ microwave detunings) compared to prior art. 
These improvements will enable higher fidelity remote entangling operations of superconducting qubits that approach the threshold for entanglement purification~\cite{Kalb-2017}.

\begin{figure}
\includegraphics[width=0.90\columnwidth]{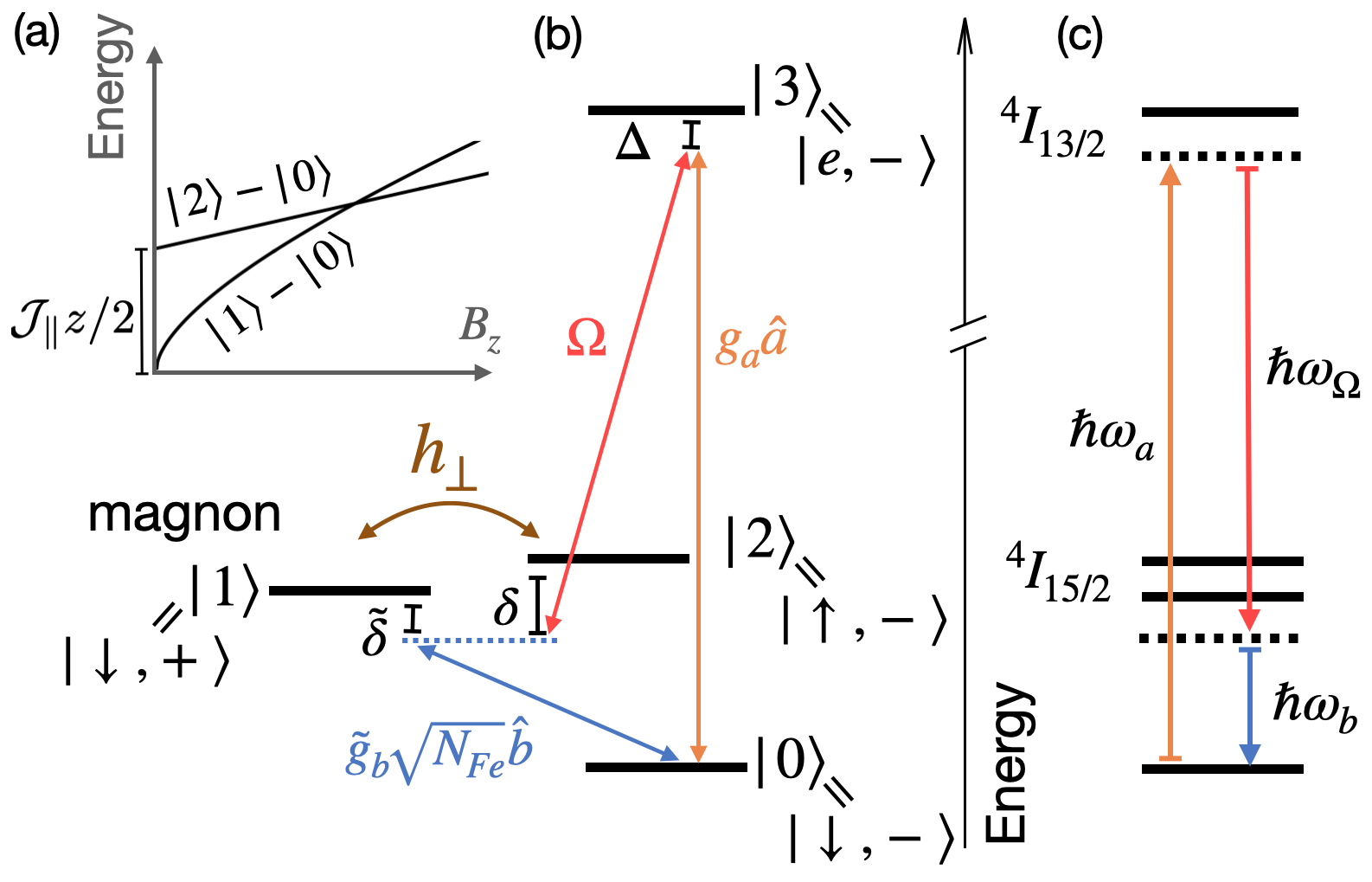}

\caption{\color{black}
(a) Crossing between the magnon excitation ($\left|0\right\rangle \rightarrow\left|1\right\rangle $) and
the erbium-spin flip ($\left|0\right\rangle \rightarrow\left|2\right\rangle $) versus $B_{z}$. 
(b) Full energy-level diagram,
including the erbium excited state ($\left|3\right\rangle$), and the transitions due to the couplings
to an optical cavity ($\text{g}_{a}\hat{a}$), to an optical pump ($\Omega$), to a magnon via
spin exchange ($h_{\perp}$), and to a magnon via the microwave cavity ($\tilde{\text{g}}_{b}\sqrt{N_{\text{Fe}}}\hat{b}$).
$\tilde{\delta}$, $\delta$, and $\Delta$ are detuning parameters
to the cavity frequencies. (c) Frequencies of the optical cavity resonance ($\omega_{a}$), microwave resonator ($\omega_{b}$), and optical pump ($\omega_{\Omega}$).
}

\label{fig: energy levels}
\end{figure}

{\color{black} The enhancement of the transduction can be derived from the linear combination of optically-active-single-spin defects interacting with the spins in a magnet.
Without loss of generality we first assume 
a single erbium ion  within
a YIG magnet }(represented by the Fe atoms)  in the presence of 
a static magnetic field, as well as
a microwave (MW) and optical (opt) fields, 
as schematically presented in Fig \ref{fig: energy levels}.

\paragraph{Hamiltonian.---}
The system above can be described with the Hamiltonian
\begin{equation}
{\cal H}\left(t\right)={\cal H}_{0}+{\cal H}_{\text{Zee}}+{\cal H}_{\text{ex}}+{\cal H}_{\text{opt}}\left(t\right)+{\cal H}_{\text{MW}}\left(t\right),
\label{eq: complete Hamiltonian}
\end{equation}
where ${\cal H}_{0}={\cal H}_{\text{Fe}}+{\cal H}_{\text{Er}}$ is
the total energy of all iron atoms and a single erbium ion in a crystal.
${\cal H}_{\text{Zee}}$ is the Zeeman term due to the presence of
an external static field $B_z$ along the $z$ direction. 
The exchange interaction ${\cal H}_{\text{ex}}$ between the single erbium spin and its $z$ nearest neighbors (n.n.) iron atoms is assumed to have the following
anisotropic form 
\begin{align}
{\cal H}_{\text{ex}} & =-\frac{1}{\hbar^{2}}\boldsymbol{S}^{\text{Er}}\cdot\boldsymbol{{\cal J}}\cdot\sum_{i\in\text{n.n.}}\boldsymbol{S}_{i}^{\text{Fe}},\nonumber \\
 & \approx-\frac{z{\cal J}_{\perp}}{2\hbar^{2}}\left(S_{+}^{\text{Er}}S_{i,-}^{\text{Fe}}+S_{-}^{\text{Er}}S_{i,+}^{\text{Fe}}\right)-\frac{z{\cal J}_{\parallel}}{\hbar^{2}}S_{z}^{\text{Er}}S_{i,z}^{\text{Fe}},
\end{align}
$\forall\;i\in\text{n.n.}$. The $\boldsymbol{S}_{i}^{\text{Fe}}$
is the magnetic moment of the $i$-th iron atom and $\boldsymbol{S}^{\text{Er}}$
the pseudospin operator of a single erbium ion.
The term with exchange coupling ${\cal J}_\parallel$ contributes to the Zeeman energy and ${\cal J}_\perp$ to the exchange of excitation between spins, i.e., converting a magnetic excitation into an erbium spin-flip transition. 
The ${\cal H}_\text{opt} (t)$ accounts for the interaction between the optical fields and the Er ion. 
The two oscillating optical fields, $\boldsymbol{E}_a (t)$ and $\boldsymbol{E}_\Omega (t)$, couple the ground and first excited manifolds ($^{4}I_{15/2}-{}^{4}I_{13/2}$) of the erbium ion through an effective electric dipole operator ($\boldsymbol{\mu}^\text{Er}$) as 
\begin{equation}
{\cal H}_{\text{opt}}\left(t\right)=-\boldsymbol{\mu}^{\text{Er}}\cdot\boldsymbol{E}_{a}\left(t\right)-\boldsymbol{\mu}^{\text{Er}}\cdot\boldsymbol{E}_{\Omega}\left(t\right).
\end{equation}
Here, $\boldsymbol{E}_{\Omega}\left(t\right) = E_\Omega \cos \left( \omega_\Omega t \right)  \hat{\boldsymbol{e}}_\Omega$ is an external pump field and  $\boldsymbol{E}_{a}\left(t\right) = E_a \cos \left( \omega_a t \right)  \hat{\boldsymbol{e}}_a$ is an upconverted optical sideband field.
In addition, the ${\cal H}_{\text{MW}} (t)$ accounts for the presence of a microwave field, $\boldsymbol{B}_{b}\left(t\right) = B_b \cos \left( \omega_b t \right) \hat{\boldsymbol{e}}_b$,
that couples to both erbium and iron spins via their magnetic dipole
moments as 
\begin{equation}
{\cal H}_{\text{MW}} \left(t\right)=\left(\mu_{B}g_{S}\hbar^{-1}\sum_{i=1}^{N_{\text{Fe}}}\boldsymbol{S}_{i}^{\text{Fe}}+\mu_{B}g_{J}\hbar^{-1}\boldsymbol{S}^{\text{Er}}\right)\cdot\boldsymbol{B}_{b}\left(t\right),
\end{equation}
where $N_\text{Fe}$ is the total number of spins in the magnet volume, and $g_S$ and $g_J$ are the g-factors of the iron spins and the erbium spins in the manifold with total angular momentum $J$, respectively. 

The Hamiltonian in Eq. (\ref{eq: complete Hamiltonian}) acts on the reduced erbium ion energy 
levels $\left|\downarrow\right\rangle \equiv\left|J=15/2,m_{S}=-1/2\right\rangle $,
$\left|\uparrow\right\rangle \equiv\left|J=15/2,m_{S}=+1/2\right\rangle $,
and $\left|e\right\rangle \equiv\left|J=13/2,m_{S}=-1/2\right\rangle $,
where $m_{S}$ is the pseudospin related to the lowest-energy doublet
within the multiplet $J$. 
It also acts on the ground and first excited states of
the magnet, which are $\left|-\right\rangle \equiv\left|\downarrow\ldots\downarrow\right\rangle $
and $\left|+\right\rangle \equiv N_{\text{Fe}}^{-1/2}\sum_{i=1}^{N_{\text{Fe}}}\left|\downarrow\ldots\uparrow_{i}\ldots\downarrow\right\rangle $.
These two states describe the uniform mode of the magnet, a magnon, in which the spins presses coherently around the $z$ axis, and makes no assumption regarding the sample geometry.
The ground state energy of the composite system is   $E_{\downarrow}\equiv\left\langle \downarrow,-\right|{\cal H}\left|\downarrow,-\right\rangle $.
An energy level diagram is illustrated in Fig.~\ref{fig: energy levels}(a).
We are particularly interested in the magnetic excitation energy $E_{m}\equiv\left\langle \downarrow,+\right|{\cal H}\left|\downarrow,+\right\rangle $
and the first spin excitation $E_{\uparrow}\equiv\left\langle \uparrow,-\right|{\cal H}\left|\uparrow,-\right\rangle $
relative to the ground state, i.e., 
\begin{align}
E_{m}-E_{\downarrow} & = \gamma\sqrt{B_{z}\left(B_{z}+M_{s}\right)} -\frac{z {\cal J}_\parallel}{2 N_\text{Fe}},
\label{eq: kittel formula} \\
E_{\uparrow}-E_{\downarrow} & = \mu_{B}g_{g}B_{z}+\frac{z{\cal J}_{\parallel}}{2},
\label{eq: Eup minus Edown}
\end{align}
where the square root in Eq. (\ref{eq: kittel formula}) follows the Kittel formula \citep{PhysRev.73.155,Kittel-8thEd-2004} for the magnetic excitation of a thin film, in which $\gamma$ is the gyromagnetic ratio and $M_S$ is the saturation magnetization. 
Here $g_g \equiv g_{J=15/2}$. 
Notably, the Ising exchange coupling ${\cal J}_{\parallel}$ acts as an effective static magnetic field significantly changing the energy of the erbium-spin transition (Eq. (\ref{eq: Eup minus Edown})), but has little effect on the magnon resonance frequency as it is normalized by the total number of spins in the magnet, 
notice ${\cal J}_{\parallel}/N_\text{Fe}$ in Eq. (\ref{eq: kittel formula}).
Finally, 
the magnon and erbium spins must be near resonance in order to allow magnon-erbium flip-flops within the secular approximation. Thus, the exchange interaction ${\cal J}_\parallel$ must be small enough to allow for the intersection of the Er and magnon resonances at an experimentally feasible magnetic field.
The following results assume a microwave frequency, set by $\boldsymbol{B}_b (t)$, to be mutually detuned from magnon and spin resonant frequencies in Eqs. (\ref{eq: kittel formula}) and (\ref{eq: Eup minus Edown}).

\paragraph{Hamiltonian in cavity QED notation.---}
First, we relabel the erbium states $\left|0\right\rangle \equiv\left|\downarrow,-\right\rangle $, $\left|2\right\rangle \equiv\left|\uparrow,-\right\rangle $,
and $\left|3\right\rangle \equiv\left|e,-\right\rangle $, and the magnon state $\left|1\right\rangle \equiv\left|\downarrow,+\right\rangle $, see Fig.~\ref{fig: energy levels}. Second,
we write the Hamiltonian in terms of transition operators $\hat{\sigma}_{i,j}\equiv\left|i\right\rangle \left\langle j\right|$,
with $i,j=0,1,2,3$. Third, we define the transition elements
\begin{align}
\left\langle 1\right|{\cal H}\left(t\right)\left|2\right\rangle  & =\hbar h_{\perp}N_{\text{Fe}}^{-1/2}\left\langle 1\right|\hat{\sigma}_{1,2}\left|2\right\rangle ,\\
\left\langle 0\right|{\cal H}\left(t\right)\left|2\right\rangle  & =\hbar\text{g}_{b}2\cos\left(\omega_{b}t\right)\left\langle 0\right|\hat{\sigma}_{0,2}\left|2\right\rangle ,\\
\left\langle 0\right|{\cal H}\left(t\right)\left|1\right\rangle  & =\hbar\tilde{\text{g}}_{b}N_{\text{Fe}}^{1/2}2\cos\left(\omega_{b}t\right)\left\langle 0\right|\hat{\sigma}_{0,1}\left|1\right\rangle ,\\
\left\langle 0\right|{\cal H}\left(t\right)\left|3\right\rangle  & =\hbar\text{g}_{a}2\cos\left(\omega_{a}t\right)\left\langle 0\right|\hat{\sigma}_{0,3}\left|3\right\rangle ,\\
\left\langle 2\right|{\cal H}\left(t\right)\left|3\right\rangle  & =\hbar\Omega2\cos\left(\omega_{\Omega}t\right)\left\langle 2\right|\hat{\sigma}_{2,3}\left|3\right\rangle,
\end{align}
related to the previous quantities as
$\text{g}_b, \tilde{\text{g}}_b \propto B_b$, $\text{g}_a \propto E_a$, $\Omega \propto E_\Omega$, and $h_\perp \propto -{\cal J}_\perp$ (more details in the SM). 
The new variables $\text{g}_b$, $\tilde{\text{g}}_b$, $\text{g}_a$, and $h_\perp$ have values of coupling per spin, and $\Omega$ is the optical pump Rabi frequency (or the pump power $\propto \Omega^2$).
From that, we build a new Hamiltonian ${\cal H}'\left(t\right)=\sum_{i,j}\left\langle i\right|{\cal H}\left(t\right)\left|j\right\rangle \left|i\right\rangle \left\langle j\right|$.
The time dependence can be removed using the rotating-wave approximation
(RWA)\citep{Walls-Milburn-2008} ($\text{e} ^{-i 2\omega t} \approx 0$, see SM), 
taking ${\cal H}'\left(t\right) \rightarrow {\cal H}_{\text{RWA}}$. 
The energy levels are then defined with respect to their detunings from
the wave frequencies, namely, we define
$\hbar\delta=\left(E_{\uparrow}-E_{\downarrow}\right)-\hbar\omega_{b}$,
$\hbar\tilde{\delta}=\left(E_{m}-E_{\downarrow}\right)-\hbar\omega_{b}$,
and $\hbar\Delta=\left(E_{e}-E_{\downarrow}\right)-\hbar\omega_{a}$,
with $\omega_{a}=\omega_{\Omega}+\omega_{b}$, as illustrated in Fig.~\ref{fig: energy levels}. The Hamiltonian
can then be written as 
%
\begin{equation}
{\cal H}_{\text{RWA}}=\hbar
\begin{pmatrix}0 & \tilde{\text{g}}_{b}N_{\text{Fe}}^{1/2} & \text{g}_{b} & \text{g}_{a} \\
\text{h.c.} & \tilde{\delta} & h_{\perp}N_{\text{Fe}}^{-1/2} & 0 \\
\text{h.c.} & \text{h.c.} & \delta & \Omega \\
\text{h.c.} & 0 & \text{h.c.} & \Delta 
\end{pmatrix}.\label{eq: Hamiltonian matrix RWA}
\end{equation}

%
Next, we suppose that the crystal is embedded in both an optical cavity
(of frequency $\omega_{a}$) and a microwave resonator (of frequency
$\omega_{b}$), such that there will be an exchange of photons between the
microwave transitions and the resonator (i.e., $\hat{\sigma}_{0,1}\rightarrow\hat{\sigma}_{0,1}\hat{b}^{\dagger}$
and $\hat{\sigma}_{0,2}\rightarrow\hat{\sigma}_{0,2}\hat{b}^{\dagger}$),
as well as the optical transition and the cavity (i.e., $\hat{\sigma}_{0,3}\rightarrow\hat{\sigma}_{0,3}\hat{a}^{\dagger}$). The optical pump field is tuned to create a three photon resonance between the microwave and optical cavity fields, i.e. $\omega_b + \omega_{\Omega}=\omega_a$. 
These transitions are schematically represented in Fig.~\ref{fig: energy levels}(c).

The Hilbert space including the cavity and the resonator becomes $\left|\psi\right\rangle =\left|\text{Er},\text{Fe}\right\rangle \left|b\right\rangle \left|a\right\rangle $,
from which we build a new Hamiltonian ${\cal H}'_\text{RWA}=\sum_{i,j}\left\langle i\right|{\cal H}_\text{RWA} \left|j\right\rangle \left|i\right\rangle \left\langle j\right|$.
Solving the eqs. of motion $i\hbar\partial_{t}\left|\psi\left(t\right)\right\rangle ={\cal H}'_\text{RWA}\left|\psi\left(t\right)\right\rangle$,
using adiabatic elimination of the higher energy states\citep{FEWELL2005125,Brion_2007,Torosov_2012} ($|\tilde{\delta}|\gg |\tilde{\text{g}}|,|h|$, and $|\delta|\gg|\Omega|,|\text{g}_b|,|h|$,
and $\left|\Delta\right|\gg|\text{g}_{a}|,|\Omega|$, with $\tilde{\text{g}} = \tilde{\text{g}}_b N^{1/2}_\text{Fe}$ and $h= h_\perp N^{-1/2}_\text{Fe}$, see SM),
we end up with an effective Hamiltonian
of the form ${\cal H}_{\text{eff}}=i\left(\kappa_{a}/2\right)\hat{a}^{\dagger}\hat{a}+i\left(\kappa_{b}/2\right)\hat{b}^{\dagger}\hat{b}+S^{*}\hat{a}^{\dagger}\hat{b}+S\hat{b}^{\dagger}\hat{a}$,
where $\kappa_{a},\kappa_{b}$ are photon leakage rates of the optical cavity and the
microwave resonator, respectively, and $S$ is the transduction rate.
{\color{black}As the pump power is a limiting constraint in practical transduction implementations,} here we focus on the {\it normalized transduction rate} $S/\Omega$, corresponding to the transduction rate normalized by the pump Rabi frequency. 
{\color{black} The linear combination of erbium-ion-spin ensemble, $N_\text{Er}$, results in}
\begin{equation}
\frac{S}{\Omega} = \frac{N_{\text{Er}}\left(\tilde{\delta}\text{g}_{b}-h_{\perp}\tilde{\text{g}}_{b}\right)\text{g}_{a}}{\Delta\left(\delta\tilde{\delta}-h_{\perp}^{2}N_{\text{Fe}}^{-1}\right)}.
\label{eq: transduction rate}
\end{equation}
Notice that the limit $h_\perp \rightarrow 0$ recovers the result in Ref.\cite{PhysRevLett.113.203601}.
Via the input-output formalism we introduce the operators $\hat{a}_{\text{in}},\hat{a}_{\text{out}},\hat{b}_{\text{in}},\hat{b}_{\text{out}}$,
from which we extract the efficiency ($\eta=2\sqrt{{\cal C}}/\left(1+{\cal C}\right)$)
and cooperativity (${\cal C}=4\left|S\right|^{2}/\left(\kappa_{a}\kappa_{b}\right)$),
see SM. 
In this scenario, the unit efficiency happens when the transduction rate is matched to the cavities leakage, i.e., the so-called impedance matching condition $2|S|=\sqrt{ \kappa_{a} \kappa_{b}}$, thus ${\cal C} = 1$ and $\eta = 1$. 
As we are going to show next, higher rates $\kappa_{a}$ and $\kappa_{b}$ are desirable to achieve higher transduction rates, mitigating the losses and leading to larger bandwidths.
Our main findings in this letter come from exploring the limit ${|h_{\perp}\tilde{\text{g}}_{b}| \gg |\tilde{\delta}\text{g}_{b}|}$ for obtaining higher transduction rates.
If we further separate the leakage into an extrinsic leakage (henceforth called coupling rate)
$\kappa_{a,c},\kappa_{b,c}$ and an intrinsic device loss $\kappa_{a,i},\kappa_{b,i}$,
the efficiency is computed as (more details in the SM)
\begin{equation}
\eta=\frac{4\left|S\right|\sqrt{\kappa_{a,c}\kappa_{b,c}}}{\left(\kappa_{a,c}+\kappa_{a,i}\right)\left(\kappa_{b,c}+\kappa_{b,i}\right)+4\left|S\right|^{2}}.
\label{eq: efficiency}
\end{equation}
Notice that $100\%$ efficiency is achieved at the impedance match condition only for zero intrinsic losses, i.e., $\kappa_{a,i}=\kappa_{b,i}=0$, however, the intrinsic losses in a device can be mitigated by a high ratio of cavity coupling to intrinsic loss $\kappa_{(a,b),c} / \kappa_{(a,b),i} \gg 1$.

\begin{figure}
\includegraphics[width=0.99\columnwidth]{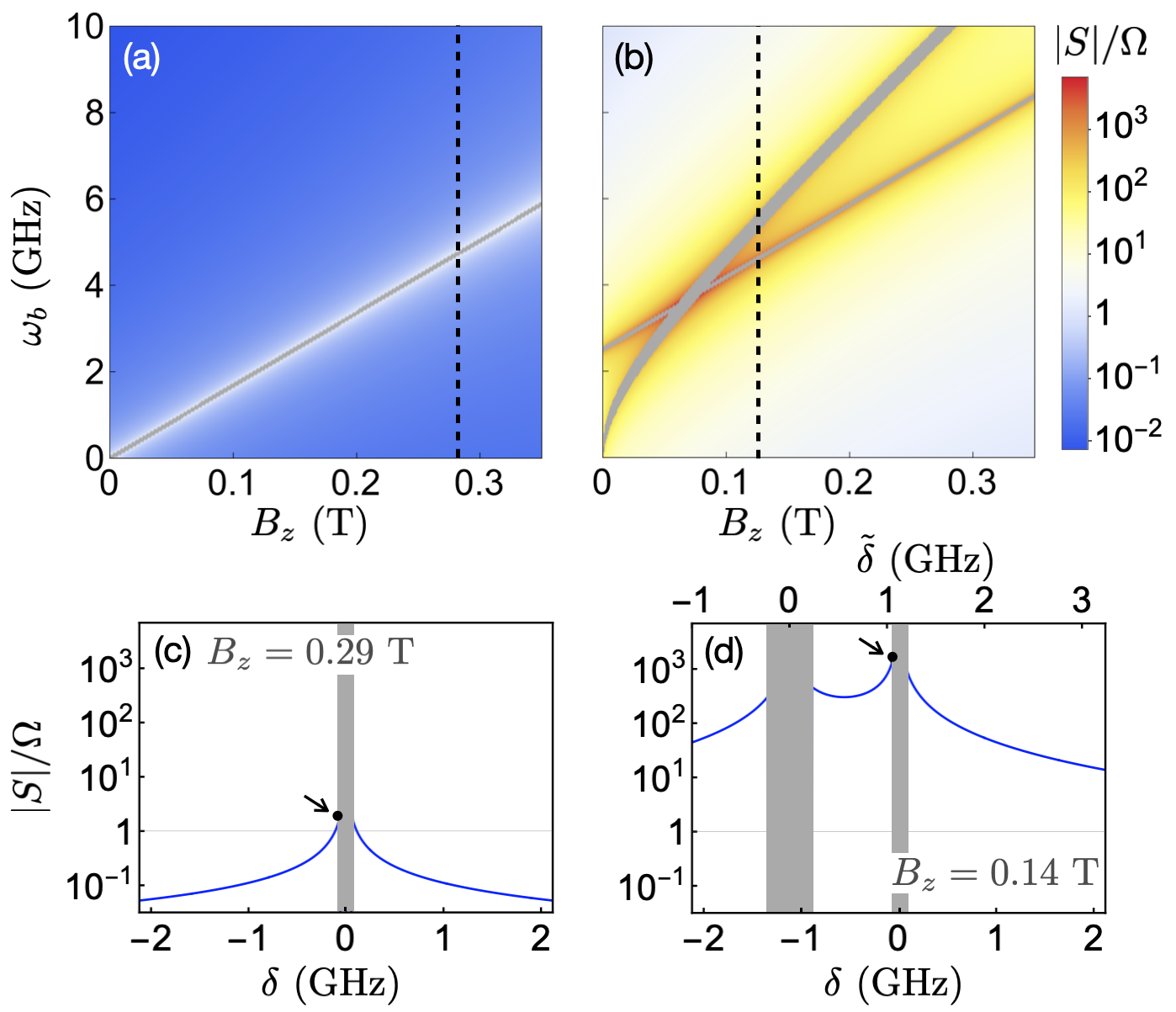}

\caption{\color{black}
Normalized transduction rate ($|S|/\Omega$) versus $B_z$ and the microwave frequency $\omega_b$, without (a) and with (b) a magnet.
The gray stripes cover the regions where the adiabatic elimination
approximation fails. Vertical-dashed lines indicate the $B_{z}$ values used in (c) and (d), which show $|S|/\Omega$ versus the detunings $\delta$ and $\tilde{\delta}$ for $\omega_b \approx 5 \text{ GHz}$ near the erbium-spin-magnon resonance.
The black points (see arrows) highlight the maximum transduction rates used in Fig.~\ref{fig: transduction rate}.
}
\label{fig: efficiency}
\end{figure}

\paragraph{Calculations.---}
In the following we show some quantitative results to illustrate magnon enhancement of rare-earth quantum transduction processes. 
We compare the normalized transduction rate ($S/\Omega$) with and without the presence of a magnet, for the idealized case of erbium-iron perpendicular interaction ${\cal J}_{\perp} = 1 \text{ THz}$ and Ising interaction ${\cal J}_{\parallel} = 1 \text{ GHz}$.
Then, we vary the erbium-iron perpendicular interaction, showing how the results connect to the case of smaller exchange anisotropy (for ${\cal J}_{\perp} \approx {\cal J}_{\parallel} \approx \text{ GHz}$).
Finally, we show how our proposal allows higher coupling rates at much lower pump power than previous proposals.

\textcolor{black}{A strong exchange interaction between erbium and iron of ${\cal J} \approx0.714\text{ THz}$ has been observed in erbium orthoferrite ($\text{ErFeO}_{3}$)\cite{doi:10.1126/science.aat5162}. 
Similarly, the spin-flop transition in $\text{Er:YIG}$ observed at $30\text{K}$ is indicative of a strong Er-Fe coupling of ${\cal J}~ \approx 0.625 \text{ THz}$\citep{doi:10.1021/acsomega.2c01334}.}
However, so far there is no evidence of anisotropy, i.e., ${\cal J}_{\perp}={\cal J}_{\parallel}={\cal J}$
in these materials. Such a strong parallel coupling shifts the spin transition away from the typical $\sim 10$GHz for superconducting qubits, as discussed in Fig.~\ref{fig: energy levels}(a).
Therefore, our proposal suggests that finding a system with Er-Fe coupling of ${\cal J}_\parallel \approx 1 \text{ GHz}$ would be ideal, while keeping ${\cal J}_\perp \approx 1 \text{ THz}$.
The strong perpendicular exchange coupling is favorable because it leads to the limit ${|h_\perp \tilde{\text{g}}_b| \gg |\tilde{\delta} \text{g}_b}|$ in Eq. (\ref{eq: transduction rate}), whereas the opposite limit recovers the non-magnetic transducer in Ref.~\cite{PhysRevLett.113.203601}.

Fig.~\ref{fig: efficiency} shows $S/\Omega$ varying with the external parameters, the resonator frequency ($\omega_b$) and the static magnetic field $B_z$, for the cases \ref{fig: efficiency}(a) without a magnet and \ref{fig: efficiency}(b) with a magnet.
For a certain set of parameters, the system reaches a resonant condition ($\delta = 0$ or $\tilde \delta = 0$) between the microwave resonator and either the spin or the magnon excitations.
As the detunings reach zero, the adiabatic elimination condition is broken, populating the excited state and causing the microwave photon to be parasitically absorbed.
Thus, gray areas are added to block regions within five linewidths of spin and magnon transitions, see SM for a list of parameters used.
Comparing figures \ref{fig: efficiency}(a) and \ref{fig: efficiency}(b), we see that $S/\Omega$ reaches orders of magnitude higher values in the presence of a magnet. 
From Figs. \ref{fig: efficiency}(c) and \ref{fig: efficiency}(d) we notice that higher rates are especially large near the crossing between the magnon and the spin transition frequencies, thus close to the adiabatic elimination condition limits. Also, the presence of a magnet allows $|S/\Omega| > 10^2$ for over $1$ GHz microwave detuning range.

\begin{figure}
\includegraphics[width=0.90\columnwidth]{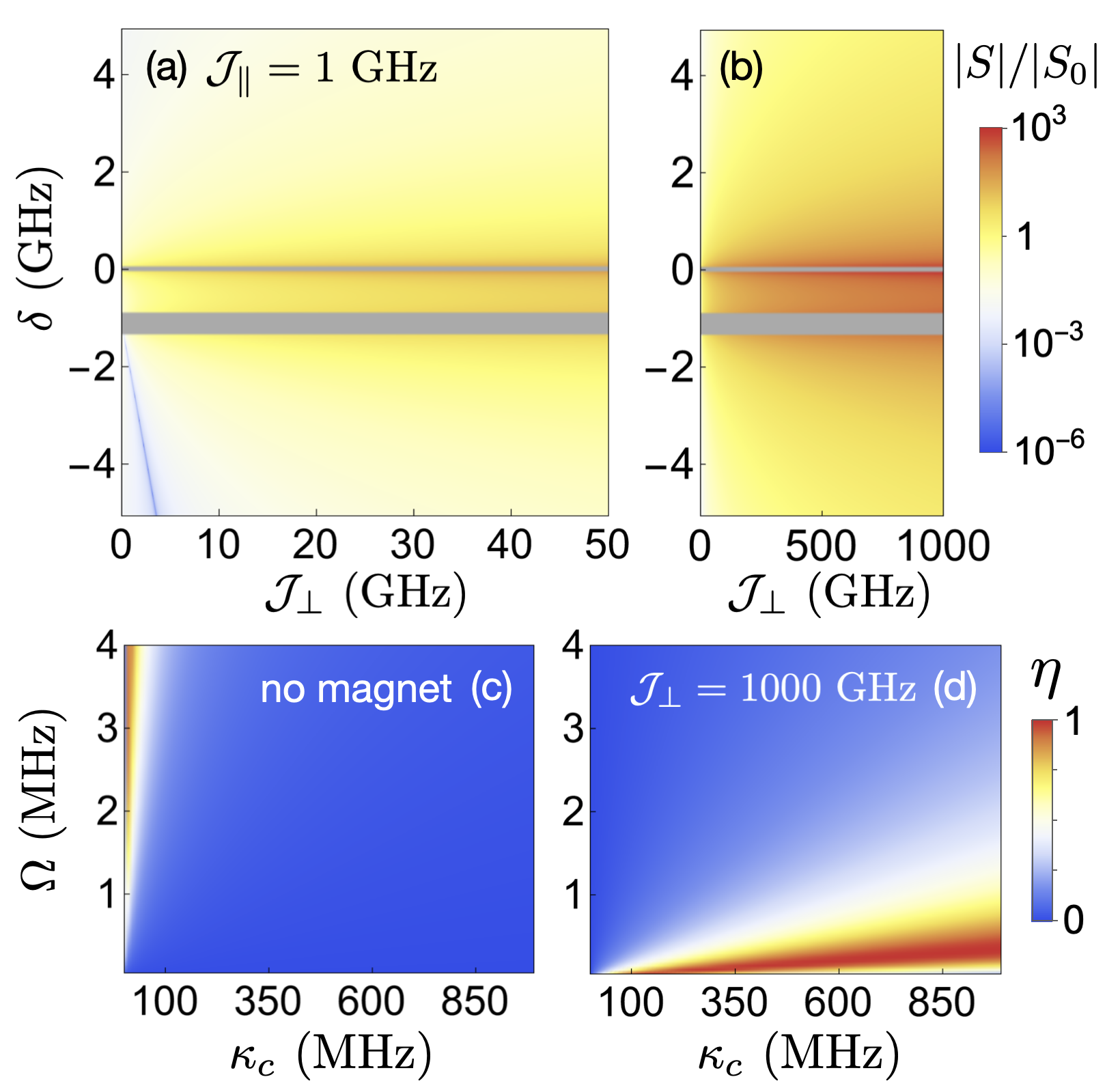}

\caption{ \color{black}
(a) Transduction rate in the presence of a magnet ($|S|$) relative
to the maximum value without a magnet $|S_{0}|$,  represented by a point in Fig.~\ref{fig: efficiency}(c), versus detuning ($\delta$) and spin-exchange interaction (${\cal J}_{\perp}$).
(b) Same as (a) over larger range of ${\cal J}_\perp$.
Gray stripes same as in Fig.~\ref{fig: efficiency}. Blue feature in (a) for
$\delta < 0$ and small ${\cal J}_{\perp}$ is due to $|S| \rightarrow 0$ for $\tilde{\delta}\text{g}_{b}\approx h_{\perp}\tilde{\text{g}}_{b}$,
see Eq. (\ref{eq: transduction rate}).
(c) and (d) Efficiency ($\eta$) versus $\Omega$ and coupling rates ($\kappa_{a,c}=\kappa_{b,c}\equiv\kappa_{c}$)
at the $|S_0|$ values marked in Figs. \ref{fig: efficiency}(c) and \ref{fig: efficiency}(d), respectively. 
}

\label{fig: transduction rate}
\end{figure}

Next, we explore the case of ${\cal J}_\perp \approx {\cal J}_\perp \approx \text{ GHz}$, that represents a magnet with weaker spin-exchange couplings, or spin-magnet dipole coupling as will be discussed below.
In Figs. \ref{fig: transduction rate}(a) and \ref{fig: transduction rate}(b) we show the transduction rate in the presence of a magnet, relative to the maximum transduction rate obtained without a magnet (defined as $|S_0|$ and marked as a black point in Fig.~\ref{fig: efficiency}(c)), as we vary the perpendicular exchange coupling.
It becomes clear that the presence of a magnet leads to higher transduction rates overall, even for lower values of ${\cal J}_\perp$, in which we see an increase of $2$ orders of magnitude at detunings close to the adiabatic elimination condition.
The feature appearing in Fig.~\ref{fig: transduction rate}(a) for negative detuning ($\delta < -1 \text{ GHz}$) and small coupling values (${\cal J}_\perp < 5 \text{ GHz}$) is due to the reduced transduction rate approaching zero for $\tilde{\delta}\text{g}_{b}\approx h_{\perp}\tilde{\text{g}}_{b}$, see Eq. (\ref{eq: transduction rate}).

To better quantify the advantages of using a magnet in the transduction process, in Figs. \ref{fig: transduction rate}(c) and \ref{fig: transduction rate}(d), we analyse the transduction efficiency, see Eq. (\ref{eq: efficiency}), as we vary the optical pump ($\Omega$) and the coupling rates ($\kappa_{a,c}=\kappa_{b,c}\equiv\kappa_{c}$).
The red region signals the near unit efficiency, where the impedance matching condition is satisfied, $2|S|=\kappa_c$.
The linear behavior in this region is the result of ${\cal J}_\perp = 1 \text{ THz}$, thus ${|h_\perp \tilde{\text{g}}_b| \gg |\tilde{\delta} \text{g}_b|}$, and the impedance matching condition leads to
\begin{align}
\frac{\kappa_{c}}{\Omega} & =\left|\frac{N_{\text{Er}}h_{\perp}\tilde{\text{g}}_{b}\text{g}_{a}}{\Delta\delta\tilde{\delta}}\right|, & \text{(with magnet)} \\
\frac{\kappa_{c}}{\Omega} & =\left|\frac{N_{\text{Er}}\text{g}_{b}\text{g}_{a}}{\Delta\delta}\right|, & \text{(without magnet)}
\end{align}
Thus, the magnet allows the maximum efficiency to be achieved at significantly lower pump powers and significantly higher cavity coupling rates.

In addition to the advantages discussed above, our proposed scheme also removes any rate penalty from mismatched optical and microwave mode volumes~\cite{Everts-2020}. This is because the microwave resonator needs to be coupled only to the iron atoms in the magnet and not the erbium ions. 
This removes a significant experimental challenge in designing co-localized optical and microwave photons, and also opens new possibilities for transduction device implementations, particularly with small mode volume \textcolor{black}{and spatially separated} optical cavities. Freedom in choice of optical cavities is particularly important as pump-induced heating \cite{Azuma:2015aa} and optical destruction of cooper pairs in superconducting microwave resonators \cite{PhysRevApplied.21.014022} have proven to be significant challenges in experimental microwave to optical transducers.

\paragraph{Discussion.---}
In the following we discuss possible routes to achieve a rare-earth magnet interaction required to realize our proposal.
It is first necessary to create an erbium spin-magnon resonance at an experimentally reasonable field. This is most easily achieved if ${\cal J}_{\parallel}\approx\text{GHz}$. 
{\color{black}
One such approach would use a dipole-dipole interaction instead of  exchange. 
For example, consider a heterostructure composed of a thin film of YIG on a substrate of Er:YAG, thus positioning the Er near to the magnet's surface and leading to strong erbium-spin-iron-spin interaction (we estimate  $\sim10\text{nm}$ for a $\sim$GHz coupling strength)\citep{PhysRev.99.1128,PhysRevB.10.787,M_G_Cottam_1971}.
}

{\color{black} A far larger transduction improvement would be achieved with large exchange anisotropy (${\cal J}_{\perp} \gg {\cal J}_{\parallel}$). 
For example, the Dzyaloshinskii-Moriya interaction (DMI) induces spin transitions as ${\left(S_{1}^{+}S_{2}^{-}-S_{1}^{-}S_{2}^{+}\right)}$, see SM. Thus, for magnets with ${\cal J}_\parallel \approx \text{GHz}$, the DMI enhances the spin transitions to fulfill our requirement ${\cal J}_\perp \gg {\cal J}_\parallel$. However, the $\text{Fe-Er}^{3+}$ DMI has not been measured.
Although the rare-earth g-tensors are known to have anisotropies that exceed $10$ in non-cubic site}~\cite{PhysRevB.74.075107,PhysRevB.77.085124}, the anisotropy of the exchange interaction itself is not well studied; 
to our knowledge, more material characterization is necessary for finding anisotropic couplings with lower strengths (${\cal J}_\parallel\approx\text{GHz}$).

Alternatively, there has been interest in using millimeter-wave qubits instead of microwave-frequency qubits for easier transduction\cite{PhysRevA.96.042305,Kumar-2023}. Our analysis directly extends to millimeter-wave magnons in which strong  ${\cal J}_{\parallel}  \sim 100\text{GHz}$ is needed to obtain resonance between the erbium spin ensemble and the magnon.

{\color{black}

We have proposed that the strong coupling between spinful optically active impurities and a magnon can implement high-efficiency transduction at rates at least two orders of magnitude faster than existing approaches based on dilute rare earth ion ensembles. 

}

\begin{acknowledgments}

This work is supported by the U. S. Department of Energy, Office of
Science: theoretical analysis of magnon mode enhancement of microwave-to-optical
transduction by BES Award Number DE-SC0023393, magnon-erbium spin
coupling by BES Award Number DE-SC0019250, supported by NQISRC Co-design Center for Quantum Advantage (C2QA) under contract number
DE-SC0012704.
\end{acknowledgments}

\bibliographystyle{apsrev4-1}
\bibliography{refs}

\appendix

\section{Supplemental Material}
In this supplemental material we give detailed description of the
Hamiltonian in different notations, as well as the applied rotating-wave
approximation. We introduce the optical and resonator modes, and reduce
the equations of motion for the entire system using the adiabatic
elimination approximation. Finally, with the input-output treatment
we are able to compute the transduction efficiency.

\section{Hamiltonian}

\subsection{Hamiltonian in solid state physics notation}

In the main text we have introduced the Hamiltonian
\begin{equation}
{\cal H}\left(t\right)={\cal H}_{0}+{\cal H}_{\text{Zee}}+{\cal H}_{\text{ex}}+{\cal H}_{\text{opt}}\left(t\right)+{\cal H}_{\text{MW}}\left(t\right),\label{eq: Hamiltonian terms}
\end{equation}
where
\begin{align}
{\cal H}_{\text{Zee}} & =\left(\mu_{B}g_{S}\hbar^{-1}\sum_{i=1}^{N_{\text{Fe}}}S_{i,z}^{\text{Fe}}+\mu_{B}g_{J}\hbar^{-1}S_{z}^{\text{Er}}\right)B_{z},\label{eq: Hamiltonian solid state notation (Zee)}\\
{\cal H}_{\text{ex}} & =-\frac{z{\cal J}_{\perp}}{2\hbar^{2}}\left(S_{+}^{\text{Er}}S_{i,-}^{\text{Fe}}+S_{-}^{\text{Er}}S_{i,+}^{\text{Fe}}\right) \nonumber \\
& \quad -\frac{z{\cal J}_{\parallel}}{\hbar^{2}}S_{z}^{\text{Er}}S_{i,z}^{\text{Fe}},\quad\forall i\in\text{n.n.} \\
{\cal H}_{\text{opt}}\left(t\right) & =-\boldsymbol{E}_{a}\left(t\right)\cdot\boldsymbol{\mu}^{\text{Er}}-\boldsymbol{E}_{\Omega}\left(t\right)\cdot\boldsymbol{\mu}^{\text{Er}},\\
{\cal H}_{\text{MW}}\left(t\right) & =\left(\mu_{B}g_{S}\hbar^{-1}\sum_{i=1}^{N_{\text{Fe}}}\boldsymbol{S}_{i}^{\text{Fe}}+\mu_{B}g_{J}\hbar^{-1}\boldsymbol{S}^{\text{Er}}\right)\cdot\boldsymbol{B}_{b}\left(t\right),\label{eq: Hamiltonian solid state notation (MW)}
\end{align}
where $\boldsymbol{S}^{\text{Er}}$ is the effective-spin operator of
a single erbium ion, the $\boldsymbol{S}_{i}^{\text{Fe}}$ is the
spin operator for the $i$-th iron atom, and $\boldsymbol{\mu}^{\text{Er}}$
is the transition-dipole-moment operator between the ground and the first excited multiplet of the
erbium ion. The $g_{S}$ is the electron's g-factor and $g_{J}$ is
the Land\' e g-factor for the erbium's crystal field level Kramer's doublets.
$N_{\text{Fe}}$ is the total number of iron atoms in the magnet.
The constants $\mu_{B}$ and $\hbar$ are the Bohr magneton and the
reduced Plank constant, respectively. In ${\cal H}_{\text{ex}}$ we
have defined constants for the spin-spin exchange coupling parallel
(${\cal J}_{\parallel}$) and perpendicular (${\cal J}_{\perp}$)
to the $\hat{\boldsymbol{z}}$ direction, and $z$ is the coordination
number, namely, the number of next neighbor iron atoms to the erbium.
The ${\cal H}_{0}$ describes both the single erbium ion energy levels
in a crystal (take for example $\text{Er}^{3+}\text{:YSO}$\citep{PhysRevB.74.075107}
or $\text{Er}^{3+}\text{:YVO}_{4}$\citep{CAPOBIANCO1997329} that
has been recently used for quantum transduction applications \citep{PhysRevA.100.033807,Rochman:2023aa})
as well as the uniform excitation of a magnet, or the Kittel mode\citep{PhysRev.73.155,Kittel-8thEd-2004}.
The Zeeman term is induced by the presence of a static field along
the $z$-direction, $B_{z}$.

The following analysis considers optical transitions between the erbium
ion's $^{4}I_{15/2}-{}^{4}I_{13/2}$ manifolds. The crystal field
split the manifold degeneracy into Kramer's pairs, and we are particularly
interested in the transitions between the lowest energy states $Z_{1}$
and $Y_{1}$ from each manifold. The presence of an external magnetic
field split the Kramer's pairs $Z_{1}^{+/-}$ and $Y_{1}^{+/-}$,
and give way to effective-spin-$1/2$ levels (principal component of the admixed wavefunctions) represented by $m_{S}=\pm1/2$.
For simplicity, we restrict the calculations to the transitions between
the following states 
\begin{align}
Y_{1}^{-}= & \left|J=13/2,m_{S}=-1/2\right\rangle \equiv\left|e\right\rangle ,\\
Z_{1}^{+}= & \left|J=15/2,m_{S}=+1/2\right\rangle \equiv\left|\uparrow\right\rangle ,\\
Z_{1}^{-}= & \left|J=15/2,m_{S}=-1/2\right\rangle \equiv\left|\downarrow\right\rangle .
\end{align}
Furthermore, the ground state ($\left|-\right\rangle $) and the uniform
excitation of the magnet ($\left|+\right\rangle $) are
\begin{equation}
\left|-\right\rangle \equiv\left|\downarrow\ldots\downarrow\right\rangle \quad\text{and}\quad\left|+\right\rangle \equiv\frac{1}{\sqrt{N_{\text{Fe}}}}\sum_{i=1}^{N_{\text{Fe}}}\left|\downarrow\ldots\uparrow_{i}\ldots\downarrow\right\rangle .
\end{equation}

Here, the optical transitions ($Z\rightarrow Y$) are treated as \underline{effective}
transition-dipole moments~\cite{PhysRev.127.750,10.1063/1.1701366,10.1063/1.1695840}, thus coupled to the electric component
of the electromagnetic waves $\boldsymbol{E}_{a}\left(t\right)$ and
$\boldsymbol{E}_{\Omega}\left(t\right)$. On the other hand, the microwave
transitions ($Z_{1}^{-}\rightarrow Z_{1}^{+}$ and $\left|-\right\rangle \rightarrow\left|+\right\rangle $)
are induced via magnetic transition-dipole moments only, thus coupled
to the magnetic component of the electromagnetic wave $\boldsymbol{B}_{b}\left(t\right)$.
Without loss of generality, for the calculations below, we have defined
\begin{align}
\boldsymbol{E}_{a}\left(t\right) & =E_{a}\cos\left(\omega_{a}t\right)\hat{\boldsymbol{e}}_{a},\qquad\boldsymbol{E}_{\Omega}\left(t\right)=E_{\Omega}\cos\left(\omega_{\Omega}t\right)\hat{\boldsymbol{e}}_{\Omega},  \\
\boldsymbol{B}_{b}\left(t\right) & =B_{b}\cos\left(\omega_{b}t\right)\hat{\boldsymbol{e}}_{b},
\end{align}
where $\hat{\boldsymbol{e}}$ are unit vectors that live in the $xy$-plane. 

If we span the Hamiltonian in eq. (\ref{eq: Hamiltonian terms}) in
our restricted basis we find the following diagonal terms
\begin{align}
\left\langle \downarrow,-\right|{\cal H}\left(t\right)\left|\downarrow,-\right\rangle  & \equiv E_{\downarrow},\qquad\left\langle \downarrow,+\right|{\cal H}\left(t\right)\left|\downarrow,+\right\rangle \equiv E_{m}\\
\left\langle \uparrow,-\right|{\cal H}\left(t\right)\left|\uparrow,-\right\rangle  & \equiv E_{\uparrow},\qquad\left\langle \uparrow,+\right|{\cal H}\left(t\right)\left|\uparrow,+\right\rangle =0,\\
\left\langle e,-\right|{\cal H}\left(t\right)\left|e,-\right\rangle  & \equiv E_{e},\qquad\left\langle e,+\right|{\cal H}\left(t\right)\left|e,+\right\rangle =0.
\end{align}
The last two terms being zero mean that we assume the systems to be
thermally initialized into the ground state, and do not populate excited
states through off-resonant driving. Anisotropy in the exchange interaction
like $J_{x}S_{x}^{\text{Er}}S_{i,x}^{\text{Fe}}+J_{y}S_{y}^{\text{Er}}S_{i,y}^{\text{Fe}}$
with $J_{x}\neq J_{y}$ would also lead to terms like $S_{+}^{\text{Er}}S_{i,+}^{\text{Fe}}+S_{-}^{\text{Er}}S_{i,-}^{\text{Fe}}$
and, therefore, non-zero values at the last two terms. The energies
defined above can be explicitly written as 
\begin{align}
E_{m}-E_{\downarrow} & =\Delta E_{m}\left(B_{z}\right)+\mu_{B}g_{S}B_{z}-\frac{{\cal J}_{\parallel}z}{2N_{\text{Fe}}},\\
E_{\uparrow}-E_{\downarrow} & =\mu_{B}g_{g}B_{z}+\frac{{\cal J}_{\parallel}z}{2},\\
E_{e}-E_{\downarrow} & =\Delta E_{e}+\mu_{B}\left(g_{e}-g_{g}\right)B_{z}.
\end{align}
The $\Delta E_{m}\left(B_{z}\right)$ adds to the linear Zeeman splitting
($\mu_{B}g_{S}B_{z}$) such that both together follow the well known
Kittel curve\citep{PhysRev.73.155,Kittel-8thEd-2004}, i.e., $\Delta E_{m}\left(B_{z}\right)+\mu_{B}g_{S}B_{z}=\gamma\sqrt{B_{z}\left(B_{z}+M_{S}\right)}$
where $\gamma$ is the gyromagnetic ratio and $M_{S}$ is the saturation
magnetization. The Land\' e g-factor for the ground and excited states
are respectively $g_{g}=g_{J=15/2}$ and $g_{e}=g_{J=13/2}$. $\Delta E_{e}$
is the erbium ion excited energy in the absent of external magnetic
field. The off-diagonal elements are
\begin{align}
\left\langle \downarrow,-\right|{\cal H}\left|\uparrow,-\right\rangle  & =\frac{\mu_{B}g_{g}}{2}\beta_{-}B_{b}\cos\left(\omega_{b}t\right),\\
\left\langle \downarrow,-\right|{\cal H}\left|\downarrow,+\right\rangle & =\frac{\mu_{B}g_{S}}{2}\sqrt{N_{\text{Fe}}}B_{b}\cos\left(\omega_{b}t\right),\label{eq: transition elements 1}\\
\left\langle \downarrow,-\right|{\cal H}\left|e,-\right\rangle  & =\mu^{\text{Er}}E_{a}\cos\left(\omega_{a}t\right), \\
\left\langle \uparrow,-\right|{\cal H}\left|e,-\right\rangle & =\mu^{\text{Er}}E_{\Omega}\cos\left(\omega_{\Omega}t\right),\\
\left\langle \downarrow,+\right|{\cal H}\left|\uparrow,-\right\rangle  & =-\frac{{\cal J}_{\perp}z\beta_{-}}{2\sqrt{N_{\text{Fe}}}},\label{eq: transition elements 3}
\end{align}
in which we have considered the same effective dipole-transition strength $\mu^{\text{Er}}$
for both spin transitions, and we have defined $\beta_{\pm}\equiv\left\langle i\right|J_{\pm}^{\text{Er}}\left|j\right\rangle $.
Except for the Hermitian conjugate partners, the elements that are
not listed in the equations above are zero.

\subsection{Hamiltonian in cavity QED notation}

In order to write the Hamiltonian above in cavity QED notation, we
conveniently relabel the erbium and iron states as
\begin{align}
\left|0\right\rangle  & \equiv\left|\downarrow,-\right\rangle ,\quad\left|1\right\rangle \equiv\left|\downarrow,+\right\rangle ,\quad\left|2\right\rangle \equiv\left|\uparrow,-\right\rangle ,\quad\left|3\right\rangle \equiv\left|e,-\right\rangle .
\end{align}
Additionally, it is usual to write the Hamiltonian in terms of operators
that represent the transitions, thus connecting different states.
We define the operators
\begin{align}
\hat{\sigma}_{i,j} & \equiv\left|i\right\rangle \left\langle j\right|,\qquad i,j=0,1,2,3.
\end{align}
From that we immediately see that the energy terms are relabelled
to 
\begin{equation}
E_{\downarrow}=E_{0},\qquad E_{m}=E_{1},\qquad E_{\uparrow}=E_{2},\qquad E_{e}=E_{3}.
\end{equation}
Using the new set of operators, we redefine the transition elements
as
\begin{align}
\left\langle \downarrow,-\right|{\cal H}\left|\uparrow,-\right\rangle  & \equiv\hbar\text{g}_{b}2\cos\left(\omega_{b}t\right)\left\langle 0\right|\hat{\sigma}_{0,2}\left|2\right\rangle , \\
\left\langle \downarrow,-\right|{\cal H}\left|\downarrow,+\right\rangle & \equiv\hbar\tilde{\text{g}}_{b}N_{\text{Fe}}^{1/2}2\cos\left(\omega_{b}t\right)\left\langle 0\right|\hat{\sigma}_{0,1}\left|1\right\rangle ,\\
\left\langle \downarrow,-\right|{\cal H}\left|e,-\right\rangle  & \equiv\hbar\text{g}_{a}2\cos\left(\omega_{a}t\right)\left\langle 0\right|\hat{\sigma}_{0,3}\left|3\right\rangle , \\
\left\langle \uparrow,-\right|{\cal H}\left|e,-\right\rangle & \equiv\hbar\Omega2\cos\left(\omega_{\Omega}t\right)\left\langle 2\right|\hat{\sigma}_{2,3}\left|3\right\rangle ,\\
\left\langle \downarrow,+\right|{\cal H}\left|\uparrow,-\right\rangle  & \equiv\hbar h_{\perp}N_{\text{Fe}}^{-1/2}\left\langle 1\right|\hat{\sigma}_{1,2}\left|2\right\rangle ,
\end{align}
where the factor of two in front of the cosine functions is for convenience
purposes only, and will simplify the Hamiltonian after the rotating-wave
approximation. The new variables $\text{g}_{b},\tilde{\text{g}}_{b},\text{g}_{a},\Omega,h_{\perp}$
can be promptly identified comparing the equations above with eqs.
(\ref{eq: transition elements 1}-\ref{eq: transition elements 3}),
the brakets of the operators $\hat{\sigma}$ above are all equal to
$1$, and are there for clarity purposes only. 
{\color{black} The $\text{g}_{b}$ and $\tilde{\text{g}}_{b}$ magnify the coupling of the microwave photon to the erbium spin transition and the magnon excitation, respectively, while $h_\perp$ magnify the coupling between erbium spin transition and magnon excitation.  
All of them represent coupling per spin.
The $\text{g}_{a}$ is the coupling between the optical photon and the erbium ground to excited multiplet transition ($Z_1^- \rightarrow Y_1$).
Finally, the $\Omega$ is related to the optical pump and quantify the Rabi oscillations between the states $Z_1^+ \rightarrow Y_1$.
}
The definitions above lead the Hamiltonian in eq. (1) to be rewritten as
\begin{align}
{\cal H}\left(t\right)= & \sum_{i=1,2,3}E_{i}\hat{\sigma}_{i,i}+\hbar h_{\perp}N_{\text{Fe}}^{-1/2}\hat{\sigma}_{1,2}\nonumber \\
 & +\hbar\text{g}_{b}2\cos\left(\omega_{b}t\right)\hat{\sigma}_{0,2}+\hbar\tilde{\text{g}}_{b}N_{\text{Fe}}^{1/2}2\cos\left(\omega_{b}t\right)\hat{\sigma}_{0,1}\nonumber \\
 & +\hbar\text{g}_{a}2\cos\left(\omega_{a}t\right)\hat{\sigma}_{0,3}+\hbar\Omega2\cos\left(\omega_{\Omega}t\right)\hat{\sigma}_{2,3}+\text{h.c.},\label{eq: Hamiltonian optics notation}
\end{align}
and we have set $E_{\downarrow}=0$ for simplicity.

\subsection{Rotating-wave approximation}

In the following procedure, we apply the rotating-wave approximation
(RWA)\citep{Walls-Milburn-2008} and end up with a non-time-dependent
Hamiltonian. We start by defining the unitary operator
\begin{equation}
R\left(t\right)=\text{e}^{i\xi t},\qquad\xi=\sum_{i=1,2,3}x_{i}\hat{\sigma}_{i,i},
\end{equation}
that brings the Hamiltonian in eq. (\ref{eq: Hamiltonian optics notation})
to the form
\begin{align}
{\cal H}_{R}\left(t\right)= & R\left(t\right){\cal H}\left(t\right)R^{\dagger}\left(t\right)-\hbar\xi\nonumber \\
= & \sum_{i=1,2,3}\left(E_{i}-\hbar x_{i}\right)\hat{\sigma}_{i,i}+\hbar h_{\perp}N_{\text{Fe}}^{-1/2}\text{e}^{i\left(x_{1}-x_{2}\right)t}\hat{\sigma}_{1,2}\nonumber \\
  +\hbar\text{g}_{b} 2\cos & \left(\omega_{b}t\right) \text{e}^{-ix_{2}t}\hat{\sigma}_{0,2}  +\hbar\tilde{\text{g}}_{b}N_{\text{Fe}}^{1/2}2\cos\left(\omega_{b}t\right)\text{e}^{-ix_{1}t}\hat{\sigma}_{0,1}\nonumber \\
+\hbar\text{g}_{a}2\cos & \left(\omega_{a}t\right)\text{e}^{-ix_{3}t}\hat{\sigma}_{0,3}+\hbar\Omega2\cos\left(\omega_{\Omega}t\right)\text{e}^{i\left(x_{2}-x_{3}\right)t}\hat{\sigma}_{2,3} \nonumber \\
 + \text{h.c.} \qquad &
\end{align}
Now we want a frequency $x_{i}$ corresponding to the external field
responsible for transitions to the energy $E_{i}$, namely, we choose
\begin{equation}
x_{1}=x_{2}=\omega_{b}\qquad\text{and}\qquad x_{3}=\omega_{a}.
\end{equation}
We further notice that these frequencies are close but not equal to
the energy levels, therefore they are detuned from each other by
\begin{equation}
\hbar\tilde{\delta}\equiv E_{1}-\hbar\omega_{b},\quad\hbar\delta\equiv E_{2}-\hbar\omega_{b},\quad\hbar\Delta\equiv E_{3}-\hbar\omega_{a}.
\end{equation}
With these definitions we bring the Hamiltonian to
\begin{align}
{\cal H}_{R}\left(t\right)= & \hbar\tilde{\delta}\hat{\sigma}_{1,1}+\hbar\delta\hat{\sigma}_{2,2}+\hbar\Delta\hat{\sigma}_{3,3}+\hbar h_{\perp}N_{\text{Fe}}^{-1/2}\hat{\sigma}_{1,2}\nonumber \\
+\hbar\text{g}_{b}2\cos & \left(\omega_{b}t\right)\text{e}^{-i\omega_{b}t}\hat{\sigma}_{0,2}+\hbar\tilde{\text{g}}_{b}N_{\text{Fe}}^{1/2}2\cos\left(\omega_{b}t\right)\text{e}^{-i\omega_{b}t}\hat{\sigma}_{0,1}\nonumber \\
+\hbar\text{g}_{a}2\cos & \left(\omega_{a}t\right)\text{e}^{-i\omega_{a}t}\hat{\sigma}_{0,3}+\hbar\Omega2\cos\left(\omega_{\Omega}t\right)\text{e}^{i\left(\omega_{b}-\omega_{a}\right)t}\hat{\sigma}_{2,3} \nonumber \\
+\text{h.c.} \qquad &
\end{align}
Through the RWA, we disregard fast oscillations which average out
over time, therefore we can approximate $1+\text{e}^{-i\omega t}\approx1$\citep{Walls-Milburn-2008}.
Using the Euler formula in the equation above and tuning the frequency
of the laser pump to be
\begin{equation}
\omega_{\Omega}=\omega_{b}-\omega_{a},
\end{equation}
we arrive at the time-independent Hamiltonian
\begin{align}
{\cal H}_{R}= & \hbar\tilde{\delta}\hat{\sigma}_{1,1}+\hbar\delta\hat{\sigma}_{2,2}+\hbar\Delta\hat{\sigma}_{3,3}+\hbar h_{\perp}N_{\text{Fe}}^{-1/2}\hat{\sigma}_{1,2}\nonumber \\
 & +\hbar\text{g}_{b}\hat{\sigma}_{0,2}+\hbar\tilde{\text{g}}_{b}N_{\text{Fe}}^{1/2}\hat{\sigma}_{0,1}\nonumber \\
 & +\hbar\text{g}_{a}\hat{\sigma}_{0,3}+\hbar\Omega\hat{\sigma}_{2,3}+\text{h.c.}\label{eq: Hamiltonian RWA}
\end{align}

\section{Optical cavity and Resonator modes}

Suppose the optical ($\omega_{a}$) and the MW ($\omega_{b}$) waves
are confined in an optical cavity and a resonator, respectively. The
cavity and resonator modes are described by the creation/annihilation
operators $\hat{a}^{\dagger}/\hat{a}$ and $\hat{b}^{\dagger}/\hat{b}$,
respectively. Here we assume that the modes' occupation are associated
to the atomic transitions such that 
\begin{equation}
\hat{\sigma}_{0,1}\rightarrow\hat{\sigma}_{0,1}\hat{b}^{\dagger},\qquad\hat{\sigma}_{0,2}\rightarrow\hat{\sigma}_{0,2}\hat{b}^{\dagger},\qquad\hat{\sigma}_{0,3}\rightarrow\hat{\sigma}_{0,3}\hat{a}^{\dagger},
\end{equation}
and their complex conjugates. The RWA Hamiltonian in eq. (\ref{eq: Hamiltonian RWA})
becomes
\begin{align}
{\cal H}_{R}= & \hbar\tilde{\delta}\hat{\sigma}_{1,1}+\hbar\delta\hat{\sigma}_{2,2}+\hbar\Delta\hat{\sigma}_{3,3}+\hbar h_{\perp}N_{\text{Fe}}^{-1/2}\hat{\sigma}_{1,2}\nonumber \\
 & +\hbar\text{g}_{b}\hat{\sigma}_{0,2}\hat{b}^{\dagger}+\hbar\tilde{\text{g}}_{b}N_{\text{Fe}}^{1/2}\hat{\sigma}_{0,1}\hat{b}^{\dagger}\nonumber \\
 & +\hbar\text{g}_{a}\hat{\sigma}_{0,3}\hat{a}^{\dagger}+\hbar\Omega\hat{\sigma}_{2,3}+\text{h.c.}
\end{align}

We restrict our basis to empty ($\left|0\right\rangle $) and singly
occupied ($\left|1\right\rangle $) cavity and resonator modes. The
entire system's basis is $\left|\psi\right\rangle =\left|\text{Er},\text{Fe}\right\rangle \left|b\right\rangle \left|a\right\rangle $.
In the matrix form, the Hamiltonian can be written as $H_{R}=\sum_{i,j=0}^{4}\left\langle i\right|{\cal H}_{R}\left|j\right\rangle \left|i\right\rangle \left\langle j\right|$,
where
\begin{align}
\left|0\right\rangle  & =\left|\downarrow,-\right\rangle \left|1\right\rangle \left|0\right\rangle ,\; \left|1\right\rangle =\left|\downarrow,+\right\rangle \left|0\right\rangle \left|0\right\rangle ,\; \left|2\right\rangle =\left|\uparrow,-\right\rangle \left|0\right\rangle \left|0\right\rangle ,\\
\left|3\right\rangle  & =\left|e,-\right\rangle \left|0\right\rangle \left|0\right\rangle ,\; \left|4\right\rangle =\left|\downarrow,-\right\rangle \left|0\right\rangle \left|1\right\rangle ,
\end{align}
therefore
\begin{equation}
H_{R}=\hbar\begin{pmatrix}0 & \tilde{\text{g}}_{b}N_{\text{Fe}}^{1/2} & \text{g}_{b} & 0 & 0\\
\text{h.c.} & \tilde{\delta} & h_{\perp}N_{\text{Fe}}^{-1/2} & 0 & 0\\
\text{h.c.} & \text{h.c.} & \delta & \Omega & 0\\
0 & 0 & \text{h.c.} & \Delta & \text{h.c.}\\
0 & 0 & 0 & \text{g}_{a} & 0
\end{pmatrix}.\label{eq: Hamiltonian matrix RWA}
\end{equation}
Notice that $H_R = {\cal H}'_\text{RWA}$ defined in the main text.

\section{Adiabatic elimination}
\label{sec: adiabatic elimination}

In this section we'll solve the equations of motion for the matrix
Hamiltonian in eq. (\ref{eq: Hamiltonian matrix RWA}), i.e., $i\hbar\partial_{t}\left|\psi\left(t\right)\right\rangle =H_{R}\left|\psi\left(t\right)\right\rangle $,
under the adiabatic elimination approximation\citep{FEWELL2005125,Brion_2007,Torosov_2012}.
For convenience, we temporarily hide the factor $\sqrt{N_{\text{Fe}}}$ by relabeling
$\tilde{\text{g}}\equiv\tilde{\text{g}}_{b}N_{\text{Fe}}^{1/2}$ and
$h\equiv h_{\perp}N_{\text{Fe}}^{-1/2}$. We'll bring that back later.
Then, we define the wavevector $\left|\psi\left(t\right)\right\rangle =\begin{pmatrix}c_{0}\left(t\right) & c_{1}\left(t\right) & c_{2}\left(t\right) & c_{3}\left(t\right) & c_{4}\left(t\right)\end{pmatrix}^{T}$,
such that the equation of motion is
\begin{equation}
i\hbar\begin{pmatrix}\dot{c}_{0}\\
\dot{c}_{1}\\
\dot{c}_{2}\\
\dot{c}_{3}\\
\dot{c}_{4}
\end{pmatrix}=\begin{pmatrix}\tilde{\text{g}}c_{1}+\text{g}_{b}c_{2}\\
\tilde{\text{g}}^{*}c_{0}+\tilde{\delta}c_{1}+hc_{2}\\
\text{g}_{b}^{*}c_{0}+hc_{1}+\delta c_{2}+\Omega c_{3}\\
\Omega^{*}c_{2}+\Delta c_{3}+\text{g}_{a}c_{4}\\
\text{g}_{a}^{*}c_{3}
\end{pmatrix}.
\end{equation}
Here, we impose that the erbium and iron are thermally initialized
into their ground states via dilution refrigerator temperatures, and
that our detunings are sufficient such that the excited states are
only virtually populated, i.e., we apply the adiabatic elimination
approximation $\dot{c}_{1}=\dot{c}_{2}=\dot{c}_{3}=0$\citep{FEWELL2005125,Brion_2007,Torosov_2012}.
This is true as long as the detunings are much larger than the coupling
strengths, i.e., $|\tilde{\delta}|\gg|\tilde{\text{g}}|,|h|$, $|\delta|\gg|\Omega|,|\text{g}_b|,|h|$,
and $\left|\Delta\right|\gg\left|\text{g}_{a}\right|,\left|\Omega\right|$.
From that, we find the following set of equations
\begin{align}
c_{3} & =-\frac{\Omega^{*}\left(h\tilde{\text{g}}^{*}-\tilde{\delta}\text{g}_{b}^{*}\right)}{\Delta\left(\delta\tilde{\delta}-h^{2}\right)-\tilde{\delta}\left|\Omega\right|^{2}}c_{0}-\frac{\left(\delta\tilde{\delta}-h^{2}\right)\text{g}_{a}}{\Delta\left(\delta\tilde{\delta}-h^{2}\right)-\tilde{\delta}\left|\Omega\right|^{2}}c_{4},\\
c_{2} & =\frac{\Delta\left(h\tilde{\text{g}}^{*}-\tilde{\delta}\text{g}_{b}^{*}\right)}{\Delta\left(\delta\tilde{\delta}-h^{2}\right)-\tilde{\delta}\left|\Omega\right|^{2}}c_{0}+\frac{\tilde{\delta}\Omega\text{g}_{a}}{\Delta\left(\delta\tilde{\delta}-h^{2}\right)-\tilde{\delta}\left|\Omega\right|^{2}}c_{4},\\
c_{1} & =\frac{\left(\left|\Omega\right|^{2}-\Delta\delta\right)\tilde{\text{g}}^{*}+\Delta h\text{g}_{b}^{*}}{\Delta\left(\delta\tilde{\delta}-h^{2}\right)-\tilde{\delta}\left|\Omega\right|^{2}}c_{0}-\frac{h\Omega\text{g}_{a}}{\Delta\left(\delta\tilde{\delta}-h^{2}\right)-\tilde{\delta}\left|\Omega\right|^{2}}c_{4},
\end{align}
and the equations of motion become 
\begin{widetext}
\begin{equation}
i\hbar\begin{pmatrix}\dot{c}_{0}\\
\dot{c}_{4}
\end{pmatrix}=\frac{1}{\Delta\left(\delta\tilde{\delta}-h^{2}\right)-\tilde{\delta}\left|\Omega\right|^{2}}\begin{pmatrix}\left[\left(\left|\Omega\right|^{2}-\Delta\delta\right)\left|\tilde{\text{g}}_{b}\right|^{2}+\Delta h\left(\tilde{\text{g}}\text{g}_{b}^{*}+\text{g}_{b}\tilde{\text{g}}^{*}\right)-\Delta\tilde{\delta}\left|\text{g}_{b}\right|^{2}\right]c_{0}-\left(h\tilde{\text{g}}-\tilde{\delta}\text{g}_{b}\right)\Omega\text{g}_{a}c_{4}\\
-\left(h\tilde{\text{g}}^{*}-\tilde{\delta}\text{g}_{b}^{*}\right)\Omega^{*}\text{g}_{a}^{*}c_{0}-\left(\delta\tilde{\delta}-h^{2}\right)\left|\text{g}_{a}\right|^{2}c_{4}
\end{pmatrix}.
\end{equation}
\end{widetext}
In the limit of $\left|\Delta\left(\delta\tilde{\delta}-h^{2}\right)\right|\gg\left|\tilde{\delta}\left|\Omega\right|^{2}\right|$,
and bringing back the $N_{\text{Fe}}$ factors, i.e., $\tilde{\text{g}}\equiv\tilde{\text{g}}_{b}N_{\text{Fe}}^{1/2}$
and $h\equiv h_{\perp}N_{\text{Fe}}^{-1/2}$, the equations above
translate into an effective Hamiltonian
\begin{equation}
{\cal H}_{\text{eff}}=\begin{pmatrix}\hat{a}^{\dagger} & \hat{b}^{\dagger}\end{pmatrix}\begin{pmatrix}\lambda_{a} & S\\
S^{*} & \lambda_{b}
\end{pmatrix}\begin{pmatrix}\hat{a}\\
\hat{b}
\end{pmatrix},\label{eq: effective Hamiltonin adiabatic elimination}
\end{equation}
with transduction rate
\begin{equation}
S\equiv\frac{\left(\tilde{\delta}\text{g}_{b}^{*}-h_{\perp}\tilde{\text{g}}_{b}^{*}\right)\Omega^{*}\text{g}_{a}^{*}}{\Delta\left(\delta\tilde{\delta}-h_{\perp}^{2}N_{\text{Fe}}^{-1}\right)}\label{eq: transduction rate per erbium}
\end{equation}
and
\begin{align}
\lambda_{a} & \equiv\frac{\left|\text{g}_{a}\right|^{2}}{\Delta},\\
\lambda_{b} & \equiv\frac{1}{\left(\delta\tilde{\delta}-h_{\perp}^{2}N_{\text{Fe}}^{-1}\right)} \Big{[}-h_{\perp}\left(\text{g}_{b}^{*}\tilde{\text{g}}_{b}+\tilde{\text{g}}_{b}^{*}\text{g}_{b}\right) \nonumber \\
& \quad -\left(\frac{\left|\Omega\right|^{2}}{\Delta}-\delta\right)\left|\tilde{\text{g}}_{b}\right|^{2}N_{\text{Fe}}+\tilde{\delta}\left|\text{g}_{b}\right|^{2}\Big{]}.
\end{align}
Notice that, in the case of non-interacting ions, the total transduction
rate is simply $N_{\text{Er}}S$, as expressed in the main text.

\section{Input-output formalism}

In this section we turn our problem into a open quantum system, in
other words, we connect the microwave resonator and the optical cavity
to the environment through the input-output formalism\citep{Walls-Milburn-2008}.
We start by evaluating the effective Hamiltonian in eq. (\ref{eq: effective Hamiltonin adiabatic elimination})
using the Heisenberg formalism, namely, $\hbar\dot{\hat{a}}=i\left[{\cal H}_{\text{eff}},\hat{a}\right]$
and $\hbar\dot{\hat{b}}=i\left[{\cal H}_{\text{eff}},\hat{b}\right]$.
It is easy to check that it can be written as
\begin{align}
\dot{\hat{a}} & =\frac{i}{\hbar}\left[{\cal H}_{S},\hat{a}\right]-\frac{\kappa_{a}}{2}\hat{a},\\
\dot{\hat{b}} & =\frac{i}{\hbar}\left[{\cal H}_{S},\hat{b}\right]-\frac{\kappa_{b}}{2}\hat{b},
\end{align}
where ${\cal H}_{S}\equiv\left(S\hat{a}^{\dagger}\hat{b}+S^{*}\hat{b}^{\dagger}\hat{a}\right)$,
$\kappa_{a}\equiv i2\hbar^{-1}\lambda_{a}$, and $\kappa_{b}\equiv i2\hbar^{-1}\lambda_{b}$.
As we can see, the $\kappa_{a},\kappa_{b}$ are extrinsic loss rates
of the cavity and resonator. The input-output formalism then introduces
new operators that represent exchange photons between the system and
the environment\citep{Walls-Milburn-2008}, in particular, we can
write
\begin{align}
\dot{\hat{a}} & =\frac{i}{\hbar}\left[{\cal H}_{S},\hat{a}\right]-\frac{\kappa_{a}}{2}\hat{a}+\sqrt{\kappa_{a}}\hat{a}_{\text{in}}=\frac{i}{\hbar}\left[{\cal H}_{S},\hat{a}\right]+\frac{\kappa_{a}}{2}\hat{a}-\sqrt{\kappa_{a}}\hat{a}_{\text{out}},\label{eq: input-output equations 1}\\
\dot{\hat{b}} & =\frac{i}{\hbar}\left[{\cal H}_{S},\hat{b}\right]-\frac{\kappa_{b}}{2}\hat{b}+\sqrt{\kappa_{b}}\hat{b}_{\text{in}}=\frac{i}{\hbar}\left[{\cal H}_{S},\hat{b}\right]+\frac{\kappa_{b}}{2}\hat{b}-\sqrt{\kappa_{b}}\hat{b}_{\text{out}},\label{eq: input-output equations 2}
\end{align}
where $\hat{a}_{\text{in}}$ is a photon coming from the environment
into the optical cavity and $\hat{a}_{\text{out}}$ is the other way
around. Similarly, the operators $\hat{b}_{\text{in}}$ and $\hat{b}_{\text{out}}$
are the exchange photons between the resonator and the environment.
To satisfy the boundary conditions at the resonator inputs and outputs,
they must obey the following relation
\begin{equation}
\hat{a}=\frac{1}{\sqrt{\kappa_{a}}}\left(\hat{a}_{\text{in}}+\hat{a}_{\text{out}}\right),\qquad\hat{b}=\frac{1}{\sqrt{\kappa_{b}}}\left(\hat{b}_{\text{in}}+\hat{b}_{\text{out}}\right).\label{eq: input-output conditions}
\end{equation}

\subsection{Solution via Fourier transformation }

In order to solve the system of equations (\ref{eq: input-output equations 1})
and (\ref{eq: input-output equations 2}) we Fourier transform the
operators
\begin{align}
\hat{a}\left(t\right) & =\frac{1}{\sqrt{2\pi}}\int_{-\infty}^{\infty}d\omega\text{e}^{i\omega t}\hat{a}\left(\omega\right),\; \hat{b}\left(t\right)=\frac{1}{\sqrt{2\pi}}\int_{-\infty}^{\infty}d\omega\text{e}^{i\omega t}\hat{b}\left(\omega\right).
\end{align}
By noticing that $\dot{\hat{a}}\left(t\right)=i\omega\hat{a}\left(t\right)$
and $\dot{\hat{b}}\left(t\right)=i\omega\hat{b}\left(t\right)$, also,
evaluating the commutators $\left[{\cal H}_{S},\hat{a}\right]=-S\hat{b}$
and $\left[{\cal H}_{S},\hat{b}\right]=-S^{*}\hat{a}$, we are able
to find 
\begin{align}
\hat{a} & =\frac{\left(-i4S\sqrt{\kappa_{b}}\hat{b}_{\text{in}}+\left(i2\omega+\kappa_{b}\right)2\sqrt{\kappa_{a}}\hat{a}_{\text{in}}\right)}{\left(i2\omega+\kappa_{a}\right)\left(i2\omega+\kappa_{b}\right)+4\left|S\right|^{2}},\\
\hat{b} & =\frac{\left(-i4S^{*}\sqrt{\kappa_{a}}\hat{a}_{\text{in}}+\left(i2\omega+\kappa_{a}\right)2\sqrt{\kappa_{b}}\hat{b}_{\text{in}}\right)}{\left(i2\omega+\kappa_{b}\right)\left(i2\omega+\kappa_{a}\right)+4\left|S\right|^{2}}.
\end{align}
Finally, using the conditions in eq. (\ref{eq: input-output conditions})
we obtain
\begin{align}
\hat{a}_{\text{out}} & =\frac{-i4S\sqrt{\kappa_{a}\kappa_{b}}}{\left(i2\omega+\kappa_{a}\right)\left(i2\omega+\kappa_{b}\right)+4\left|S\right|^{2}}\hat{b}_{\text{in}} \nonumber \\
& \quad +\frac{-\left(i2\omega-\kappa_{a}\right)\left(i2\omega+\kappa_{b}\right)-4\left|S\right|^{2}}{\left(i2\omega+\kappa_{a}\right)\left(i2\omega+\kappa_{b}\right)+4\left|S\right|^{2}}\hat{a}_{\text{in}},\\
\hat{b}_{\text{out}} & =\frac{-i4S^{*}\sqrt{\kappa_{b}\kappa_{a}}}{\left(i2\omega+\kappa_{b}\right)\left(i2\omega+\kappa_{a}\right)+4\left|S\right|^{2}}\hat{a}_{\text{in}} \nonumber \\
& \quad +\frac{-\left(i2\omega-\kappa_{b}\right)\left(i2\omega+\kappa_{a}\right)-4\left|S\right|^{2}}{\left(i2\omega+\kappa_{b}\right)\left(i2\omega+\kappa_{a}\right)+4\left|S\right|^{2}}\hat{b}_{\text{in}}.
\end{align}
The efficiency of the transduction is given by the first coefficient
$\hat{b}_{\text{in}}$, that converts an input optical wave $\hat{a}_{\text{out}}$
into an output microwave $\hat{b}_{\text{in}}$. For constant resonator
mode occupations, such that $\dot{\hat{a}}=\dot{\hat{b}}=0$ thus
$\omega=0$, we find the efficiency to be
\begin{equation}
\eta=\frac{4\left|S\right|\sqrt{\kappa_{a}\kappa_{b}}}{\left(\kappa_{a}\kappa_{b}+4\left|S\right|^{2}\right)}.
\end{equation}
The perfect impedance match condition that leads to a maximum efficiency
($\eta=1$) is obtained for $2\left|S\right|=\sqrt{\kappa_{a}\kappa_{b}}$.
It is common to define the cooperativity factor ${\cal C}\equiv4\left|S\right|^{2}/\left(\kappa_{a}\kappa_{b}\right)$,
that in this case should be the closest to unit as possible, and can
be related to the efficiency through $\eta=2\sqrt{{\cal C}}/\left(1+{\cal C}\right)$.

\subsection{Solution including losses }

In the previous section, the input-output formalism was used to include
external couplings to the cavity and the resonator. Those couplings
are via extrinsic (or intentional) losses, i.e., the photons are not
lost to the environment. However, we can include intrinsic (or unintentional)
losses by rewriting Eqs. (\ref{eq: input-output equations 1}) and
(\ref{eq: input-output equations 2}) as
\begin{align}
\dot{\hat{a}} & =-iS\hat{b}-\frac{\left(\kappa_{a,c}+\kappa_{a,i}\right)}{2}\hat{a}+\sqrt{\kappa_{a,c}}\hat{a}_{\text{in}}+\sqrt{\kappa_{a,i}}\hat{a}_{\text{in,loss}},\\
\dot{\hat{a}} & =-iS\hat{b}+\frac{\left(\kappa_{a,c}+\kappa_{a,i}\right)}{2}\hat{a}-\sqrt{\kappa_{a,c}}\hat{a}_{\text{out}}-\sqrt{\kappa_{a,i}}\hat{a}_{\text{out,loss}},\\
\dot{\hat{b}} & =-iS^{*}\hat{a}-\frac{\left(\kappa_{b,c}+\kappa_{b,i}\right)}{2}\hat{b}+\sqrt{\kappa_{b,c}}\hat{b}_{\text{in}}+\sqrt{\kappa_{b,i}}\hat{b}_{\text{in,loss}},\\
\dot{\hat{b}} & =-iS^{*}\hat{a}+\frac{\left(\kappa_{b,c}+\kappa_{b,i}\right)}{2}\hat{b}-\sqrt{\kappa_{b,c}}\hat{b}_{\text{out}}-\sqrt{\kappa_{b,i}}\hat{b}_{\text{out,loss}}.
\end{align}
where $\kappa_{a,c}$ and $\kappa_{b,c}$ are the extrinsic loss rates,
while $\kappa_{a,i}$ and $\kappa_{b,i}$ are the intrinsic loss rates.
Similar calculations to the previous section lead to the following
efficiency
\begin{equation}
\eta=\frac{4\left|S\right|\sqrt{\kappa_{a,c}\kappa_{b,c}}}{\left(\kappa_{a,c}+\kappa_{a,i}\right)\left(\kappa_{b,c}+\kappa_{b,i}\right)+4\left|S\right|^{2}}.\label{eq: efficiency}
\end{equation}
The efficiency now is limited by the intrinsic losses and reaches
it maximum value when $2\left|S\right|=\sqrt{\kappa_{a}\kappa_{b}}$
such that
\begin{equation}
\eta_{\text{max}}=\frac{2}{\left(1+\kappa_{a,i}/\kappa_{a,c}\right)\left(1+\kappa_{b,i}/\kappa_{b,c}\right)+1}.
\end{equation}
Notice that we have included cavity and resonator losses only, and
other types of losses such as undesired ion decay process or magnon
damping were not considered.

\section{Calculations Parameters}

In this section we estimate the parameters used in the calculations of the transduction rate, Eq. (\ref{eq: transduction rate per erbium}), as well as the efficiency, Eq. (\ref{eq: efficiency}).
For the transducer without a magnet, hereon called Case 1, we based our choices on Refs. [\onlinecite{PhysRevA.100.033807,PhysRevB.103.214305}] and the parameters are summarized in Tables \ref{tbl: microwave parameters}-\ref{tbl: sample parameters}.
It considers a 3-dimensional microwave resonator containing as $\text{Er:YSO}$ sample in it.
For the transducer in the presence of a magnet, hereon called Case 2, there are additional parameters,
given in Tables \ref{tbl: microwave parameters with magnetic host}-\ref{tbl: other parameters}. 
For the magnet-resonator coupling we based the parameters on Ref. [\onlinecite{doi:10.1063/1.4904857}], which considers a YIG-film on a split-ring resonator, although there is no transduction involved.

In Table \ref{tbl: sample parameters}, the volume ($V$) was estimated
to be the transduction active region (i.e., the mode volume of the
optical or the microwave cavity), and calculated as the following: $V_{1}=AL=13.6\text{mm}^{3}$,
with $L=12\text{mm}$ and $A=\pi0.6^{2}\text{mm}^{2}$\citep{PhysRevA.100.033807}. 
The authors mention that the experiment counts with $1.28 \times 10^{15}$ active erbium ions.

In Case 2, we considered an YIG volume of $V_{3}=\left(1.5\times0.8\times0.025\right)\text{mm}^{3}=0.03\text{mm}^{3}$, thus assumed to be the transduction mode volume.  
The number of Fe atoms in was estimated by acknowledging
that the unit-cell volume of YIG is $1981.37 \text{\AA}^{3}$ with $40$
Fe atoms\citep{doi:10.1063/1.5018795}, therefore, a sample volume of
$0.03\text{ mm}^{3}$ has a total of $6\times10^{17}$ atoms. 
The volume above is smaller than the sample in Ref. [\onlinecite{doi:10.1063/1.4904857}], however, we may say that a larger magnet only affects the limits of the adiabatic elimination condition $|\tilde{\delta}| \gg |\tilde{\text{g}}_b| N_\text{Fe}^{1/2}$, see Section \ref{sec: adiabatic elimination} above.
Additionally, Ref. [\onlinecite{doi:10.1063/1.4904857}] provides the magnet-resonator coupling $|\tilde{\text{g}}_b|$ as in Table \ref{tbl: microwave parameters with magnetic host}.
The exchange coupling between the erbium ion and the iron spins were also taken from the literature, see Table \ref{tbl: other parameters}. However, although we consider parameters from Er:YIG in this example, there is no indication of the anisotropic behavior ${\cal J}_\parallel =10^{-3} {\cal J}_\perp$ in this material; 
workarounds to this issue were discussed in the main text.

\begin{widetext}

\begin{table}[H]
\hfill{}%
\begin{tabular}{|c|c|c|c|c|c|}
\hline 
Ref. & $\text{g}_{b}$(kHz) & $\kappa_{b,c}$(MHz) & $\kappa_{b,i}$(MHz) & $\sigma_{b}$(MHz) & $\omega_{b}$(GHz)\tabularnewline
\hline 
\hline 
Cases 1 and 2\citep{PhysRevA.100.033807,PhysRevB.103.214305} & $0.001$ & $0.75^{*}$ & $0.717$ & $3$ & $5$\tabularnewline
\hline 
\end{tabular}\hfill{}
\caption{Microwave frequency parameters: $\text{g}_{b}=$ coupling strength
per erbium ion, $\kappa_{b,c}=$ cavity coupling rate, $\kappa_{b,i}=$ cavity intrinsic loss rate, $\sigma_{b}=$ inhomogeneous broadening linewidth, and $\omega_{b}=$ cavity resonance frequency. ($^{*}$)
These are presented as free parameters in the results and are responsible for the high transduction rates. }
\label{tbl: microwave parameters}
\end{table}

\begin{table}[H]
\hfill{}%
\begin{tabular}{|c|c|c|c|c|c|c|}
\hline 
Ref. & $\text{g}_{a}$(kHz) & $\kappa_{a,c}$(MHz) & $\kappa_{a,i}$(MHz) & $\sigma_{a}$(MHz) & $\omega_{a}$(THz) & $\Omega$(MHz)\tabularnewline
\hline 
\hline 
Cases 1 and 2\citep{PhysRevA.100.033807,PhysRevB.103.214305} & $0.052$ & $7.95$ & $1.7$ & $150$ & $195$ & $11.5^{\dagger}$\tabularnewline
\hline 
\end{tabular}\hfill{}
\caption{Optical frequency parameters: $\text{g}_{a}=$ coupling strength
per erbium ion, $\kappa_{a,c}=$ cavity coupling rate, $\kappa_{a,i}=$ cavity intrinsic loss rate,
$\sigma_{a}=$ inhomogeneous broadening
linewidth, $\omega_{a}=$ cavity resonance frequency, and $\Omega=$ Rabi frequency. 
($^{\dagger}$) Assumed value.}
\label{tbl: optical parameters}
\end{table}

\begin{table}[H]
\hfill{}%
\begin{tabular}{|c|c|c|c|c|c|c|}
\hline 
Ref. & $V$(mm$^{3}$) & $N_{\text{Er}}$ & $N_{\text{Fe}}$ & $\text{g}_{a,\text{tot}}$(GHz) & $\text{g}_{b,\text{tot}}$(MHz) & $T$(K)\tabularnewline
\hline 
\hline 
Case 1\citep{PhysRevA.100.033807,PhysRevB.103.214305} & $13.6$ & $1.28\text{e}15$ & - & $1.9$ & $37$ & $0.04$\tabularnewline
\hline 
Case 2\citep{doi:10.1063/1.4904857} & $0.03$ & $1.28\text{e}15$ & $6\text{e}37$ & $1.9$ & $17$ & $295$\tabularnewline
\hline 
\end{tabular}\hfill{}
\caption{Geometric parameters: $V$ is the volume of the transduction active
region in Case 1 and volume of the magnet in Case 2, $\text{g}_{a,\text{tot}}=\text{g}_{a}\sqrt{N_{\text{Er}}}$ is the total
optical coupling strength to the erbium ions, and $\text{g}_{b,\text{tot}}=\text{g}_{b}\sqrt{N_{\text{Er}}}$
is the total microwave coupling strength to the erbium ions.}
\label{tbl: sample parameters}
\end{table}
\end{widetext}

\begin{table}[H]
\hfill{}%
\begin{tabular}{|c|c|c|c|c|}
\hline 
Ref. & $\tilde{\text{g}}_{b}$(kHz) & $\tilde{\text{g}}_{b,\text{tot}}$(MHz) & $\tilde{\sigma}_{b}$(MHz) & $\omega_{b}$(GHz)\tabularnewline
\hline 
\hline 
Case 2\citep{doi:10.1063/1.4904857} & $5.8\text{e-}5$ & $45$ & $1.4$ & $5$\tabularnewline
\hline 
\end{tabular}\hfill{}

\caption{Microwave frequencies in the presence of a magnetic host: $\tilde{\text{g}}_{b}=$
coupling strength per iron atom, $\tilde{\text{g}}_{b,\text{tot}}=\tilde{\text{g}}_{b}\sqrt{N_{\text{Fe}}}=$
total microwave coupling strength to the iron atoms, $\tilde{\sigma}_{b}=$ inhomogeneous broadening
linewidth, and $\omega_{b}=$ cavity resonance frequency. }

\label{tbl: microwave parameters with magnetic host}
\end{table}

\begin{table}[H]
\hfill{}%
\begin{tabular}{|c|c||c|c|}
\hline 
Parameter & Value & Parameter & Value\tabularnewline
\hline 
\hline 
$g_{S}$ & $2$ & $z$ & $5$\tabularnewline
\hline 
$g_{g}$ & $1.2$ & ${\cal J}_{\perp}$~\cite{doi:10.1126/science.aat5162,doi:10.1021/acsomega.2c01334} & $0.714$ THz\tabularnewline
\hline 
$g_{e}$ & $1.1$ & ${\cal J}_{\parallel}$ & $10^{-3}{\cal J}_{\perp}$\tabularnewline
\hline 
$\beta_{-}$ & $7.9$ & \multirow{1}{*}{---} & ---\tabularnewline
\hline 
\end{tabular}\hfill{}

\caption{The Er-Fe coupling ${\cal J}_{\parallel}={\cal J}_{\perp}=2.95\text{ meV}\approx0.714\text{ THz}$
was taken from Ref. \citep{doi:10.1126/science.aat5162} for erbium
orthoferrite $\text{ErFeO}_{3}$. That value agrees with the $30\text{K}$
spin flop transition found in the $\text{Er:YIG}$\citep{doi:10.1021/acsomega.2c01334}.
{\color{black} For a large transduction enhancement, we deliberately assumed ${\cal J}_{\parallel} \neq {\cal J}_\perp$.} }

\label{tbl: other parameters}
\end{table}

\section{Dzyaloshinskii-Moriya interaction}

Consider the following interacting Hamiltonian between two spins
\begin{align}
{\cal H}_{\text{I}} & =\boldsymbol{S}_{1}\cdot\boldsymbol{{\cal J}}\cdot\boldsymbol{S}_{2}\nonumber \\
 & =\boldsymbol{S}_{1}\cdot\boldsymbol{{\cal J}}_{S}\cdot\boldsymbol{S}_{2}+\boldsymbol{S}_{1}\cdot\boldsymbol{{\cal J}}_{A}\cdot\boldsymbol{S}_{2},
\end{align}
where $\boldsymbol{{\cal J}}=\boldsymbol{{\cal J}}_{S}+\boldsymbol{{\cal J}}_{A}$
is decomposed by its symmetric and anti-symmetric contributions. It
is possible to show that the anti-symmetric part is the DM interaction,
\begin{equation}
\boldsymbol{S}_{1}\cdot\boldsymbol{{\cal J}}_{A}\cdot\boldsymbol{S}_{2}=\boldsymbol{D}\cdot\left(\boldsymbol{S}_{1}\times\boldsymbol{S}_{2}\right)\equiv{\cal H}_{\text{DM}},
\end{equation}
where $\boldsymbol{D}$ is determined by the symmetries of the neighboring
sites. The symmetric contribution is accounted in the main text. 

Using the spin components and defining $\boldsymbol{D}\equiv\left(D_{x},D_{y},D_{z}\right)$,
the DM Hamiltonian can be written as 
\begin{align}
{\cal H}_{\text{DM}}= & \frac{D_{x}}{2i}\left[\left(S_{1}^{+}-S_{1}^{-}\right)S_{2}^{z}-S_{1}^{z}\left(S_{2}^{+}-S_{2}^{-}\right)\right]\nonumber \\
 & +\frac{D_{y}}{2}\left[S_{1}^{z}\left(S_{2}^{+}+S_{2}^{-}\right)-\left(S_{1}^{+}+S_{1}^{-}\right)S_{2}^{z}\right]\nonumber \\
 & -\frac{D_{z}}{2i}\left[S_{1}^{+}S_{2}^{-}-S_{1}^{-}S_{2}^{+}\right].
\end{align}
Usually the $z$-direction is defined along the magnetization, i.e.,
$\boldsymbol{M}\parallel\hat{\boldsymbol{z}}$. Here we are interested
in the case of aligned magnetization and $\boldsymbol{D}$ such that
\begin{equation}
\boldsymbol{M}\parallel\boldsymbol{D}\parallel\hat{\boldsymbol{z}},
\end{equation}
in which the DMI contributes for spin transitions only, i.e., 
\begin{equation}
{\cal H}_{\text{DM}}=-\frac{D_{z}}{2i}\left[S_{1}^{+}S_{2}^{-}-S_{1}^{-}S_{2}^{+}\right].
\end{equation}

In Ref. \citep{PhysRevLett.117.037203} they found a $\left|\boldsymbol{D}\right|=338\text{ MHz}$
for interacting $\text{Nd}^{3+}\text{-}\text{Nd}^{3+}$ pairs in $\text{YVO}_{4}$.
The DMI between $\text{Fe-Er}^{3+}$ has not been demonstrated.

\end{document}